\documentclass[pre]{revtex4}
\usepackage{graphicx}
\usepackage{dcolumn}
\usepackage{bm}
\usepackage{amsmath,amssymb}

\newcommand{\grad}{\nabla}

\newcommand {\pdd}[2]{\frac{\partial #1}{\partial #2}}

\newcommand {\vel}{\mathbf{u}}

\begin{document}

\title{The Formation of Lake Stars}

\author{Victor C. Tsai$^{1}$}
\email{vtsai@fas.harvard.edu}
\author{J. S.~Wettlaufer$^2$}
\email{john.wettlaufer@yale.edu}
\affiliation{$^1$Department of Earth \& Planetary Sciences, Harvard University, Cambridge, Massachusetts,  02138}
\affiliation{$^2$Departments of Geology \& Geophysics and Physics, Yale University\\
 New Haven, Connecticut 06520-8109}

\begin{abstract}
Star patterns, reminiscent of a wide range of diffusively controlled growth forms from snowflakes to Saffman-Taylor fingers, are ubiquitous features of ice covered lakes.  Despite the commonality and beauty of these ``lake stars'' the underlying physical processes that produce them have not been explained in a coherent theoretical framework.  Here we describe a simple mathematical model that captures the principal features of lake-star formation; radial fingers of (relatively warm) water-rich regions grow from a central source and evolve through a competition between thermal and porous media flow effects in a saturated snow layer covering the lake.  The number of star arms emerges from a stability analysis of this competition and the qualitative features of this meter-scale natural phenomena are captured in laboratory experiments. 
\end{abstract}

\date{11 April, 2007}

\pacs{45.70.Qj,	82.40.Ck, 92.40.qj, 92.40.Vq}
\maketitle

\begin{figure}[!b]
\centering
\includegraphics[width=6.5in,trim = 0 0 0 0,clip]{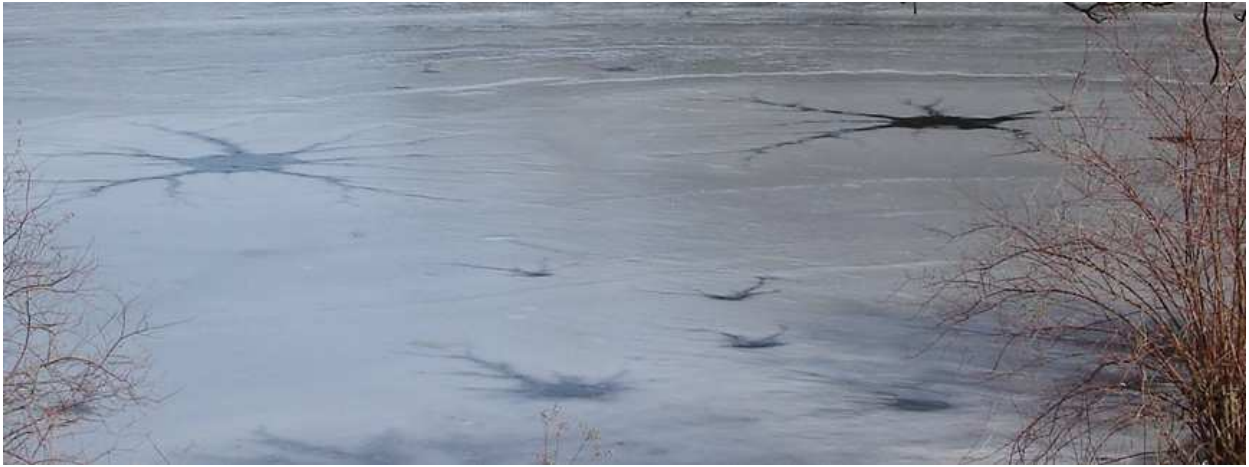}
\caption{Typical lake star patterns.  The branched arms are approximately 1 m in length. Quonnipaug Lake, Guilford, Connecticut, 8 March, 2006.}
\label{fig:photo}
\end{figure}

\section{Introduction}

The scientific study of the problems of growth and form occupies an anomalously broad set of disciplines.  Whether the emergent patterns are physical or biological in origin, their quantitative description presents many challenging and compelling issues in, for example, applied mathematics 
\cite{Shelley}, biophysics \cite{Brenner}, condensed matter \cite{Cross} and geophysics \cite{Goldenfeld} wherein the motion of free boundaries is of central interest.   In all such settings a principal goal is to predict the evolution of a boundary that is often under the influence of an instability.  Here we study a novel variant of such a situation that occurs naturally on the frozen surfaces of  lakes.

Lakes commonly freeze during a snowfall.  When a hole forms in the ice cover, relatively warm lake water will flow through it and hence through the snow layer.  In the process of flowing through and melting the snow this warm water creates dark regions.  The pattern so produced looks star-like (see Figure \ref{fig:photo}) and we refer to it as a ``lake star''.  These compelling features have been described qualitatively a number of times (e.g.~\cite{Knight,Katsaros,Woodcock}) but work on the formation process itself has been solely heuristic.  Knight~\cite{Knight} outlines a number of the physical ideas relevant to the process, but does not translate them into a predictive framework to model field observations.  Knight's main idea is that locations with faster flow rates melt preferentially, leading to even faster flow rates and therefore to an instability that results in fingers.  This idea has features that resemble those of many other instabilities such as, for example, those observed during the growth of binary alloys \cite{Worster}, in flow of water through a rigid hot porous media \cite{Woods}, or in more complex geomorphological settings \cite{Schorghofer}, and we structure our model accordingly.

Katsaros~\cite{Katsaros} and Woodcock~\cite{Woodcock} attribute the holes from which the stars emanate and the patterns themselves to thermal convection patterns within the lake, but do not measure or calculate their nature.  However, often the holes do not exhibit a characteristic distance between them but rather form from protrusions (e.g. sticks that poke through the ice surface)~\cite{Knight} and stars follow thereby ruling out a convective mechanism as being necessary to explain the phenomena.  The paucity of literature on this topic provides little more than speculation regarding the puncturing mechanism  but lake stars are observed in all of these circumstances.  Therefore, while hole formation is necessary for lake star formation, its origin does not control the mechanism of pattern formation, which is the focus of the present work. 

\section{Theory}

 The water level in the hole is higher than that in the wet snow--slush--layer \cite{Knight} and hence  we treat this warm water \footnote{A finite body of fresh water cooled from above will  have a maximum below ice temperature below  of 4 $^\circ$C.} region as having a constant height above the ice or equivalently a constant pressure head, which drives flow of water through the slush layer, which we treat as a Darcy flow of water at $0^\circ$C.  We model the temperature field within the liquid region with an advection-diffusion equation and impose an appropriate ({\em Stefan}) condition for energy conservation at the water-slush interface.  The water is everywhere incompressible.  Finally, the model is closed with an outer boundary condition at which the pressure head is assumed known.  
 
 Although we lack {\em in-situ} pressure measurements, circular water-saturated regions (a few meters in radius) are observed around the lake stars.  Hence, we assume that the differential pressure head falls to zero somewhere in the vicinity of this circular boundary.  The actual boundary at which the differential pressure head is zero is not likely to be completely uniform (as in Figure 4 of Knight~\cite{Knight}) but treating it as uniform is a good approximation in the linear regime of our analysis.  Finally, we treat the flow as two-dimensional.  Thus, although the water in direct contact with ice must be at $0^\circ$C,  we consider the depth-averaged temperature, which is above freezing.  Additionally, the decreasing pressure head in the radial direction must be accompanied by a corresponding drop in water level.  Therefore, although the driving force is more accurately described as deriving from an axisymmetric gravity current, the front whose stability we assess is controlled by the same essential physical processes that we model herein.  Our analysis could be extended to account for these three-dimensional effects.
 
\begin{figure}
\begin{center}
\includegraphics[width=3in,trim = 0 0 0 0,clip]{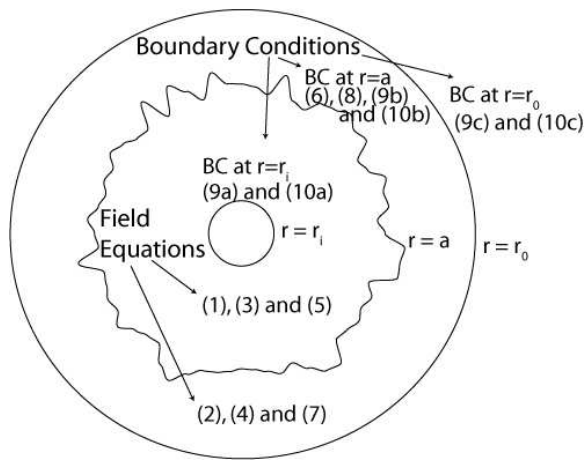}
\caption{\label{fig:schematic} Schematic of the geometry of the model. The perspective is looking down on a nascent star.  The equations (refer to text for numbering) are shown in the domains of the system where they are applicable.}
\end{center}
\end{figure}

The system is characterized by the temperature $T$, a Darcy fluid velocity $\vel$, pressure $p$, and an evolving liquid-slush interface $a$.  The liquid properties are $\kappa$ (thermal diffusivity), $C_P$ (specific heat at constant pressure) and $\mu$ (dynamic viscosity) and the slush properties are $\Pi$ (permeability), $\xi$ (solid fraction) and $L$ (latent heat).  We non-dimensionalized the equations of motion by scaling the length, temperature, pressure and velocity with $r_0$,  $T_0$, $p_0$, and $ \frac{\Pi p_0}{\mu r_0}$, respectively.  Thus, our model consists of the following system of dimensionless equations:

\begin{eqnarray}
\label {temp1n}
\pdd{\theta}{t} + \vel \cdot \grad \theta = \epsilon \grad^2 \theta & r_0 < r < a(\phi,t), \\
\label {temp2n}
\theta = 0 & a(\phi,t) < r < 1,
\end{eqnarray}

\begin{eqnarray}
\label{pres1n}
p = 1 & r_i < r < a(\phi,t), \\
\label{pres2n}
\grad^2 p = 0 & a(\phi,t) < r < 1,
\end{eqnarray}

\begin{eqnarray}
\label{continuityn}
\grad \cdot \vel = 0 & r_i < r < a(\phi,t), \\
\label{u_matchn}
\vel \mid_{a_-} = \vel \mid_{a_+} & r = a(\phi,t), \\
\label{darcyn}
\vel = -\grad p & a(\phi,t) < r < 1,
\end{eqnarray}
with boundary conditions

\begin{eqnarray}
\label{stefann}
\dot{a} = -\frac{\epsilon}{S} \grad \theta \quad r = a(\phi,t), \\
\label{temp_bcn}
\theta =  \begin{cases}
1 & r = r_i \\
0 & r = a(\phi,t) \\
0 & r = 1
\end{cases}, 
\end{eqnarray}
and
\begin{eqnarray}
\label{pres_bcn}
p = \begin{cases}
1 & r = r_i \\
1 & r = a(\phi,t) \\
0 & r = 1
\end{cases},
\end{eqnarray}
where (\ref{temp1n}) describes the temperature evolution in the liquid, (\ref{pres2n}) and (\ref {continuityn}) describe mass conservation with a Darcy flow (\ref{darcyn}) in the slush, (\ref{stefann}) is the Stefan condition, and (\ref{temp_bcn}) and (\ref{pres_bcn}) are the temperature and pressure boundary conditions, respectively  (see Figure \ref{fig:schematic}).  Note that  (\ref{pres1n}) and (\ref{continuityn}) can both be satisfied since the liquid region has an effectively infinite permeability.  

The dimensionless parameters $\epsilon$ and $S$ of the system are given by
\begin{equation}
\epsilon \equiv \frac{\kappa}{u_0 r_0}, \qquad \mbox{and} \qquad
S \equiv \frac{\xi L}{C_P T_0}, 
\end{equation}
which describe an inverse Peclet number and a Stefan number respectively.  Because the liquid must be less than or equal to $4^\circ$C,  we make the conservative estimates that $T_0 < 4^\circ$C, $\xi > 0.3$, and use the fact that $L/C_P \approx 80^\circ$C  from which we see $S > 6 \gg 1$.  Using $\kappa \approx 10^{-7} \mathrm{m^2 s^{-1}}$, and the field observations of Knight~\cite{Knight} to constrain  $u_0$ ($1 \mathrm{cm/hr} < u_0 < 10 \mathrm{cm/hr}$) and $r_0$ ($0.3 \mathrm{m} < r_0 < 3 \mathrm{m}$), we find that $\epsilon < 0.1 \ll 1$.  We therefore employ the quasi-stationary ($S \gg 1$) and large Peclet number ($\epsilon \ll 1$) approximations, and hence equations (\ref {temp1n}) - (\ref {pres_bcn}) are easily solved for a purely radial flow with cylindrical symmetry (no $\phi$ dependence) and circular liquid-slush interface.  This (boundary layer) solution is

\begin{equation}
\label {blv}
\vel = u \hat{r} = -\frac{1}{\ln(a_0)} \frac{1}{r} \hat{r} \quad r_i < r < 1,
\end{equation}

\begin{equation}
\label {blp}
p_b = \frac{\ln(r)}{\ln(a_0)} \quad r > a_0,
\end{equation}

\begin{equation}
\label {blt}
\theta_0 = 1 - \left(\frac{r}{a_0}\right)^{\frac{1}{\epsilon} (-1/\ln(a_0) + 2\epsilon)} \quad r < a_0,
\end{equation}

\begin{equation}
\label {bla}
\frac{S a_0 \dot{a}_0}{-1/\ln(a_0)+2\epsilon} = 1,
\end{equation}
where equation (\ref {bla}) has an approximate implicit solution for $a_0$ given by
\begin{equation}
\label {bla2}
\frac{a_0^2}{4} - \frac{1}{2} a_0^2 \ln(a_0) = \frac{t}{S}.
\end{equation}

We perform a linear stability analysis around this quasi-steady cylindrically symmetrical flow.  Proceeding in the usual way, we allow for scaled perturbations in $\theta$ and $a$ with scaled wavenumber $k' = \epsilon k$, non-dimensional growth rate $\sigma$, and amplitudes $f(r)$ and $g$ respectively.  Keeping only terms linear in $\epsilon$, $1/S$ and $g$, we solve (\ref {pres2n}) subject to (\ref {pres_bcn}), substitute into (\ref {u_matchn}) and satisfy (\ref {continuityn}) and (\ref {temp1n}).  This gives the non-dimensional growth rate ($\sigma$) as a function of scaled wave number ($k'$): 

\begin{equation}
\label{stability}
\sigma = \frac{1}{2a_0 \ln^2(a_0) S}\left(\sqrt{1 + 4 k'^2 \ln^2(a_0)} - 1\right)
\left(\frac{a_0}{-k' \ln(a_0)} - 1\right).
\end{equation}

Equation (\ref{stability}) can be approximated in $0 \le x \lesssim 1$ as
\begin{equation}
\label{stability_2}
\sigma \approx \frac{a_0}{\ln^2(a_0) S} x(1-x),
\end{equation}
where $x \equiv -k' \ln(a_0)/a_0$.

\begin{figure}
\begin{center}
\includegraphics[width=3in,trim = 0 0 0 0,clip]{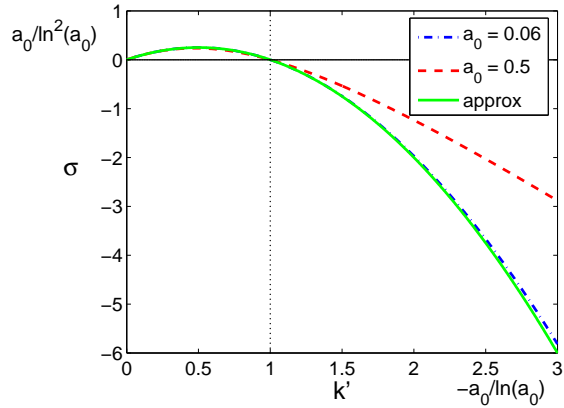}
\caption{\label{fig:stability_curve} Stability curve: Non-dimensional growth rate $\sigma$ versus non-dimensional wavenumber $k'$.  Scales for the axes are given at the upper left ($\sigma$ axis) and the lower right corners ($k'$ axis).  $\sigma$ is plotted for the range of plausible $a_0$ (dot-dashed blue and dashed red curves) and for the approximation (\ref{stability_2}) (solid green curve).}
\end{center}
\end{figure}

The stability curve (\ref{stability}) and the approximation (\ref {stability_2}) are plotted in Figure \ref{fig:stability_curve}.  The essential features of (\ref {stability}) are a maximum in the range $0 < k' < a_0/\ln(a_0)$, zero growth rate at $k' = a_0/\ln(a_0)$ and a linear increase in stability with $k'$ for large $k'$.  The long-wavelength cut-off is typical of systems with a Peclet number, here with the added effect of latent heat embodied in the Stefan number.  This demonstrates the competition between the advection and diffusion of heat and momentum (in a harmonic pressure field); the former driving the instability and the latter limiting its extent.  
The maximum growth rate occurs at approximately
\begin{equation}
\label{stability_kap}
k'_{max} \approx \frac{a_0}{-2 \ln(a_0)},
\end{equation}
with (non-dimensional) growth rate
\begin{equation}
\label{stability_sap}
\sigma_{max} \approx \frac{a_0}{4 S \ln^2(a_0)}.
\end{equation}

Translating (\ref {stability_kap}) and (\ref {stability_sap}) back into dimensional quantities, we find that the most unstable mode has angular size given by
\begin{equation}
\label{stability_deg}
\phi_{degrees} = \frac{720^\circ \kappa}{u_0 r_0} \left(\frac{r_0}{a_0}\right) \ln\left(\frac{r_0}{a_0}\right),
\end{equation}
and has growth rate given by
\begin{equation}
\label{stability_ap}
\sigma_{dim} = \frac{u_0}{4 S r_0 \ln^2(r_0/a_0)}\left(\frac{a_0}{r_0}\right).
\end{equation}

\section{Extracting information from field observations}

\begin{figure}
\begin{center}
\includegraphics[width=3in,trim = 0 0 0 0,clip]{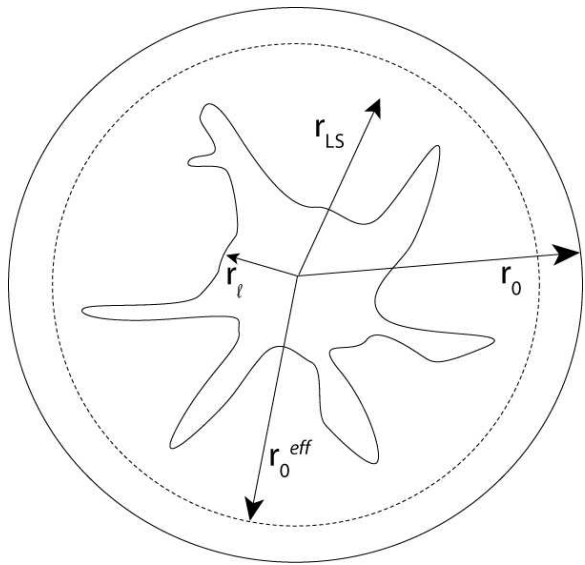}
\caption{\label{fig:obs_schem} Schematic showing $r_0$, $r_0^{eff}$,  
$r_{{LS}}$ and $r_{\ell}$.}
\end{center}
\end{figure}

Field observations of lake stars cannot be {\em controlled}.  A  
reasonable estmate for $r_0$ is the radius of the wetted (snow)  
region around the lake stars, and observations~\cite
{Knight,Woodcock,Katsaros} bound the value as $1.5 \mathrm{m}  
\lesssim r_0 \lesssim 4 \mathrm{m}$.  This is simply because if there  
were significant excess pressure at this point then the wetting front  
would have advanced further.   However, it is also possible that the  
effective value of $r_0$, say $r_0^{eff}$, is less than this either  
because the wetted radius is smaller earlier in the star formation  
process or because the ambient pressure level is reached at smaller  
radii. Here, we take $a_0$  to be the radius of the roughly circular  
liquid-filled region at the center of the lake star ($r_{\ell}$) as  
the best approximation during the initial stages of star formation  
(see Figure \ref{fig:obs_schem}).
Field observations show that $0.1 \mathrm{m} \lesssim r_{\ell}  
\lesssim 0.5 \mathrm{m}$,
~\cite{Knight,Woodcock,Katsaros} and hence $0.07 \lesssim r_{\ell}/
r_0 \lesssim 0.15$.  We note that equations (\ref
{stability_deg}) and (\ref {stability_ap}) are more sensitive to $a_0/
r_0$ than $a_0$ or $r_0$ independently\footnote{For the later stages 
of growth, clearly in the nonlinear regime not treated presently,  
$a_0$ may also be interpreted as the radius of the lake star ($r_{LS}
$).  Field observations show $1 \mathrm{m} \lesssim r_{{LS}} \lesssim  
2 \mathrm{m}$ ~\cite{Knight,Woodcock,Katsaros} and hence $0.3  
\lesssim r_{{LS}}/r_0 \lesssim 0.6$.}.  With this interpretation of  
$r_0$ we find a reasonable estimate of $u_0$ as $1.4 \cdot 10^{-5}  
\mathrm{m/s} \lesssim u_0 \lesssim 2.8 \cdot 10^{-5} \mathrm{m/s}$.  
Using these parameter values, the most unstable mode should have  
wavelength between $8^\circ$ and $130^\circ$.  Letting the number of  
branches be $N = 360^\circ/\phi_{deg}$, then $3 < N < 45$ and we clearly encompass the  
observed values for lake stars ($4 < N < 15$), but note that values  
($N>15$) are never seen in the field.

Despite the dearth of field observations, many qualitative features  
embolden our interpretation.
For example, the stars with larger values of $a_0/r_0$ have a larger  
number of branches.  Moreover, for any value of $a_0/r_0$, our  
analysis predicts an increase in $N$ with $r_0$ and $u_0$.  Indeed,  
$u_0$ increases with  $p_0$ (higher water height within the slush  
layer) and  $\Pi$ (less well-packed snow).  Therefore, we ascribe  
some of the variability among field observations to variations in  
these quantities (which have not been measured in the field) and the remainder to nonlinear effects.
Because the dendritic arms are observed long after onset and  
are far from small perturbations to a radially symmetric pattern, as  
one might see in the initial stages of the Saffman-Taylor  
instability, the process involves non-linear cooperative phenomena.  
Hence,  our model should only approximately agree with observations.  
Although a rigorous non-linear analysis of the long term star  
evolution process
(e.g.~\cite{Cross}) may more closely mirror field  
observations, the present state of the latter does not warrant that  
level of detail.  Instead, we examine the model  
physics through simple proof of concept experimentation described presently. 

\section{Demonstrating Lake Stars in the Laboratory}

\begin{figure}
\begin{center}
\includegraphics[width=3in,trim = 0 0 0 0,clip]{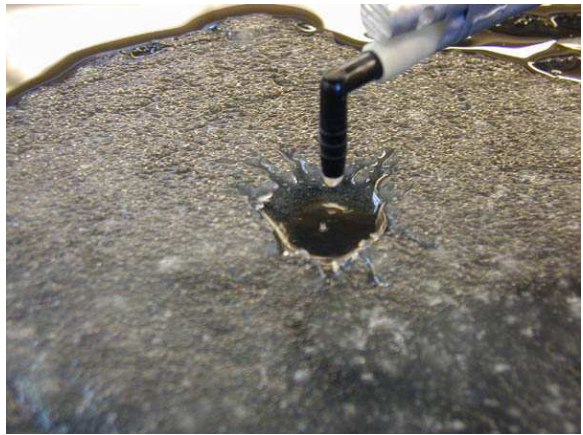}
\caption{\label{fig:fingers} Typical experimental run where small-
scale fingers are present.  For scale, the nozzle head has diameter  
of 5 mm.}
\end{center}
\end{figure}

\begin{figure}
\begin{center}
\includegraphics[width=3in,trim = 0 0 0 0,clip]{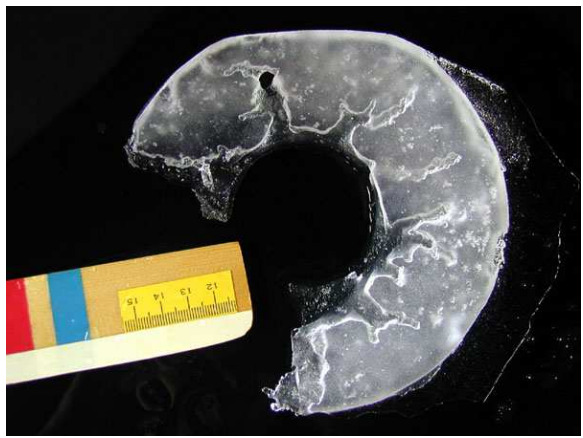}
\caption{\label{fig:channels} Typical run where channels form.  This  
picture is taken from the underside.  Note: part of the slush broke  
off when it was flipped to image it.  The ruler scale is in
cm.}
\end{center}
\end{figure}

A 30 cm diameter circular plate is maintained below freezing ($\approx -0.5^\circ$C),  and on  
top of this we place a 0.5 to 1 cm deep layer of slush through which  
we flow $1^\circ$C water.  Given the technical difficulties  
associated with its production, the grain size, and hence the  
permeability, of the slush layer, is not a controlled variable.    
This fact influences our results quantitatively. In fourteen runs we varied the initial size of the  
water-filled central hole ($a_0$), that of the circular slush layer  
($r_0$), and the flow rate ($Q$), which determines $u_0$.  The flow  
rate is adjusted manually so that the water level ($h_0$) in the  
central hole remains constant \footnote{In many of the runs, we begin  
the experiment without the central hole.  In practice, however, the  
first few drops of warm water create a circular hole with radius one  
to three times the radius of the water nozzle ($0.5 \mathrm {cm} <a_0<1.0 \mathrm {cm}$).  It is significantly more difficult to  
prepare a uniform permeability sample with a circular hole initially  
present; these runs are therefore more difficult to interpret.}.
Fingering is observed in every experimental run and hence we conclude  
that fingers are a robust feature of the system.    Two distinct  
types of fingering are observed: small-scale fingering (see Figure  
\ref{fig:fingers}) that forms early in an experimental run, and  
larger channel-like fingers (see Figure \ref{fig:channels}) that are  
ubiquitous at later times and often extend from the central hole to  
the outer edge of the slush.  Since the channel-like fingers provide  
a direct path for water to flow, effectively {\em shorting} Darcy  
flow within the slush, their subsequent dynamics are not directly  
analogous to those in natural  lake stars.  However, in {\em all}  
runs, the initial small-scale fingers have the characteristics of  
lake stars and hence we focus upon them.  We note that because the  
larger channel-like fingers emerge out of small-scale fingers, they  
likely represent the non-linear growth of the linear modes of  
instability, a topic left for future study.  Finally, we measure 
the  distance between fingers ($d_f$), so that for each experiment we can
calculate $u_0 = Q/(2 \pi r_0 h_0)$, $\phi_{calc} \equiv \phi_
{degrees}$, from equation (\ref {stability_deg}), and $\phi_{obs} =  
180^\circ d_f / (\pi a_0)$, and we can thereby  compare experiment,  
theory and field observations.

\begin{figure}
\begin{center}
\includegraphics[width=4in,trim = 0 0 0 0,clip]{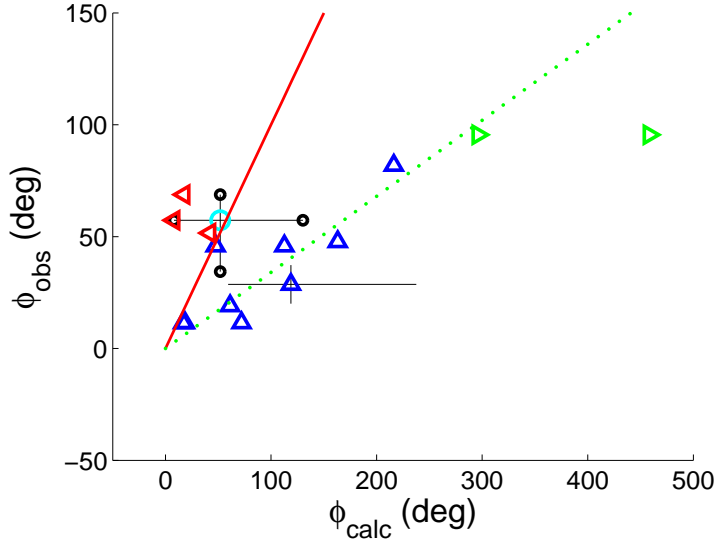}
\caption{\label{fig:comp} Comparison of theory, experiment and field observations.  Circles are field observations (cyan = best constrained field observation, black = range of plausible field observations), triangles are experimental results (blue upward-pointing triangles were unambiguous; red left-pointing triangles have channels but show no clear small-scale fingers, so channel spacing is taken for $d_f$; green right-pointing triangles were compromised by the quality of the images).  Errors are approximately 0.3 cm, 0.5 cm, 2 mm, 5 ml/min and 0.2 cm (respectively) for the five measured quantities.  All experimental results thus have error bars of at least a factor of two in the x-coordinate and 30\% in the y-coordinate.  Typical error bars are shown on one measurement.  The solid red line is the theoretical prediction; the dotted green line is the best fit line to the blue triangles.}
\end{center}
\end{figure}

In Figure \ref{fig:comp} we plot $\phi_{obs}$ versus $\phi_{calc}$ for the various field observations for which we have estimates of parameters, the laboratory experiments described above, and the model [equation (\ref {stability_deg})].  There is a large amount of scatter in both the experimental and observational data and the data does not lie on the one-to-one curve predicted by the model.  However, the experiments are meant to demonstrate the features of the model predictions, and the results have the correct qualitative trend (having a best-fit slope of 0.34).  We also attempt to find trends in the experimental data not represented by the model by comparing $y \equiv \phi_{obs}/\phi_{calc}$ vs. various combinations of control parameters ($\equiv x$) including $r_0$, $a_0$, $r_0/a_0$, $r_0 u_0$, $r_0/a_0 \ln(r_0/a_0)$ and $\ln(r_0/a_0)/(a_0 u_0)$.  For all plots of $y$ vs. $x$, our model predicts a zero slope (and y-intercept of 1).  A non-random dependence of $y$ on $x$ would point to failure of some part of our model.  Thus, to test the validity of our model, we perform significance tests on all non-flagged data with the null hypothesis being a non-zero slope. In all cases, the null hypothesis is accepted (not rejected) at the 95\% confidence level.  Thus, although the agreement is far from perfect, the simple model captures all of the significant trends in the experimental data.

\section {Conclusions}
\label{conclusion}

By generalizing and quantifying the heuristic ideas of Knight~\cite{Knight}, we have constructed a theory that is able to explain the radiating finger-like patterns on lake ice that we call lake stars.  The model yields a prediction for the wavelength of the most unstable mode as a function of various physical parameters that agrees with field observations.  Proof of concept experiments revealed the robustness of the fingering pattern, and to leading order the results also agree with the model.  There is substantial scatter in the data, and the overall comparison between field observations, model and experiment demonstrates the need for a comprehensive measurement program and a fully nonlinear theory which will yield better quantitative comparisons.  However, the general predictions of our theory capture the leading order features of the system. 

\section {Acknowledgements}
\label{acknowledge}

We thank K. Bradley and J. A. Whitehead for laboratory and facilities support.  This research, which began at the Geophysical Fluid Dynamics summer program at the Woods Hole Oceanographic Institution, was partially funded by National Science Foundation (NSF) grant OCE0325296, NSF Graduate Fellowship (VCT), NSF grant OPP0440841 (JSW), and Department of Energy grant DE-FG02-05ER15741 (JSW).

\end{document}